# Mixing due to Solution-switch Limits the Performance of Electro-sorption for Desalination

Weifan Liu[a], Longqian Xu[a], Zezhou Yang[a], Xudong Zhang[a], Shihong Lin[a, b]

[a] Department of Civil and Environmental Engineering, Vanderbilt University, Nashville, Tennessee 37235-1831, USA

[b] Department of Chemical and Bimolecular Engineering, Vanderbilt University, Nashville, Tennessee 37235-1831, USA

# Abstract

Electro-sorption (ES) is a research frontier in electrochemical separation, with proven potential applications in desalination, wastewater treatment, and selective resource extraction. However, due to the limited adsorption capacity of film electrodes, ES requires short circuiting or circuit reversal, accompanied by solution-switch between the feed solution and receiving solution, to sustain desalination over many charging-discharge cycles. In the literature, solution-switch have been commonly ignored to simplify experimental procedures and its impacts on separation performance are thus not well understood. This study aims to provide a quantitative analysis of the impacts of mixing due to solution-switch on the performance of ES-based desalination. A numerical model of ES has been employed to evaluate the adverse effects of solution-switch on the desalination performance in three different operation modes. The analysis reveals that the impacts of mixing due to solution-switch are more severe with a larger concentration difference between the desalinated water and the brine and provides insights on the effectiveness of increasing electrode loading or specific capacity in mitigating the detrimental impacts of mixing. Even with state-of-the-art systems, producing freshwater from seawater or even brackish water with medium-to-high salinity is practically challenging due to the presence of solution-switch.

## Introduction

Electrochemical separation stands as a burgeoning frontier in ion and molecular separation due to its free of chemical use, operational flexibility, scalability, and wide-spectrum applicability.[1] Electro-sorption (ES) or capacitive deionization (CDI) is an important category of electrochemical separation inspired by the mechanism of battery and supercapacitor for energy storage.[1–3] Extensive research has been performed to develop ES for desalination,[1,4] water treatment,[5] nutrient recovery,[6,7] and resource extraction.[8] ES, employing capacitive or Faradaic electrodes, adsorbs ions from aqueous solutions during charging and releases them upon voltage polarity reversal during discharge.[9] Recent years have also seen a surge of interest in applications of ES beyond desalination,[10] notably in direct lithium extraction (DLE) and nutrient recovery.[8,11,12]

The architectures of ES cells are predominantly based on film electrodes. These architectures include flow-by ES,[2,13] flow-through ES,[13,14] inverted ES,[15] and those integrating ion exchange membranes (IEMs), such as membrane ES,[16,17] rocking chair ES,[18] and hybrid ES.[19] Despite the difference in specific configurations, ES based on film electrodes share a similar general working principle, i.e., ions are adsorbed from the feed solution to the electrodes in the charging half-cycle, stored in the electrode until saturation (for constant voltage operation) or the cell voltage becomes too high (for constant current operation), and released to a receiving solution in the discharge half-cycle.[1,9,19]

The ion storage sites are micropores in activated carbon (AC) electrodes[19] and crystal lattice in intercalation electrodes.[20] The space between particles of AC or intercalation materials is called macropores which are typically saturated with the solution contacting the electrodes to promote fast ion transport (Figure 1a). All film-electrode-based ES configurations require physically switching the feed and the receiving solutions between the charging and discharge half-cycles, rendering the ES operation non-continuous.[21,22] A complete ES cycle includes the charging (adsorption) and discharge (desorption) half-cycles, as well as two solution-switch steps between the two half-cycles (Figure 1b). These solution-switch steps impose critical limitations on the performance of ES.[23] These limitations are likely recognized in the research community but have never been systematically investigated and quantified.

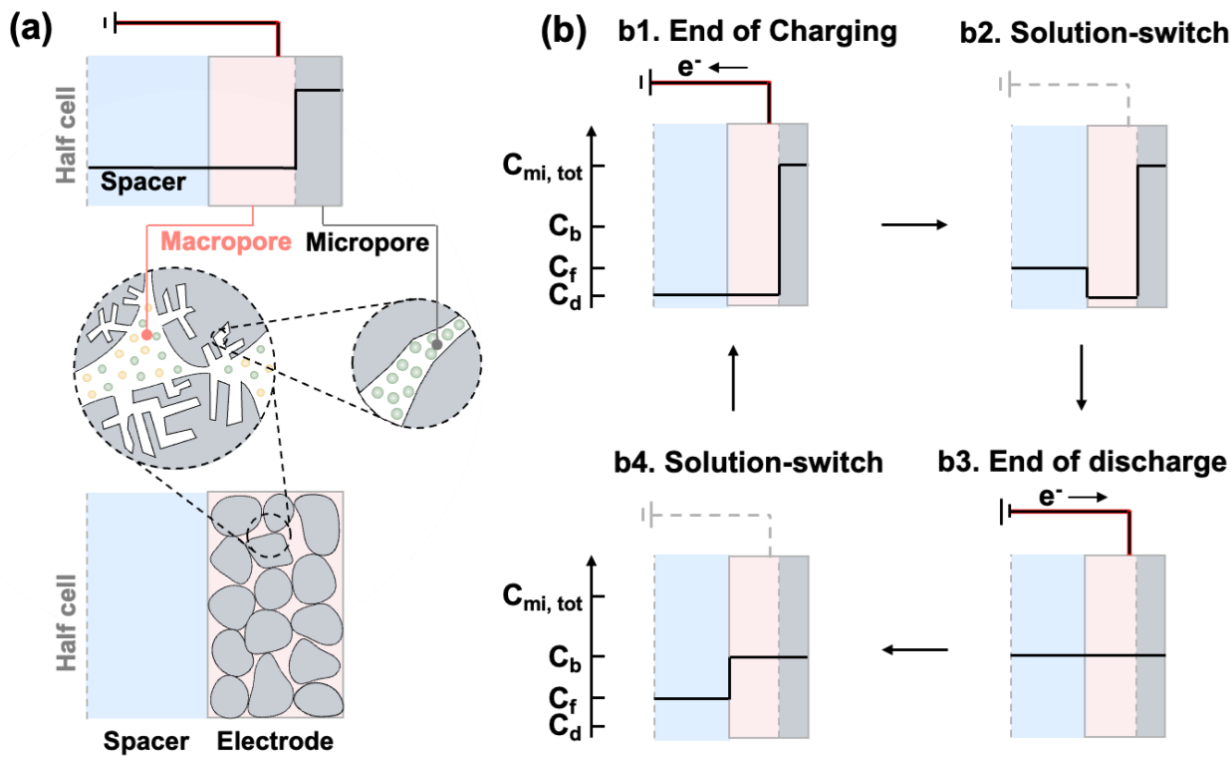

**Figure 1.** (a) Schematics of an ES half-cell comprising the spacer channel (i.e., flow channel) and the AC electrode with both macropores and micropores. The macropores and micropores occupy the same macroscopic electrode domain as in the bottom schematic. However, the top schematic intentionally split the ES half-cell into three distinct domains to facilitate the semi-quantitative representation of salt or ion concentrations in these three phases. Such as representation is used throughout Figure 1b. (b) Semi-quantitative concentration profiles for different stages of a full charging-discharge cycle, including (b1) end of charging, (b2) solution-switch after charging and prior to discharge, (b3) end of discharge, and (b4) solution-switch after discharge and prior to charging. The black lines provide a semi-quantitative representation of the concentration profiles with $C_f$, $C_b$, and $C_d$ representing the salt concentrations of the feed, brine, and diluate streams, respectively. $C_{mi,tol}$ is the total ion concentration in the micropores.

High-performance ES electrodes typically require excellent hydrophilicity and high porosity to facilitate fast ion transport between the solution and ion storage sites of the electrodes.[24,25] Consequently, solution-switch in film electrode-based ES systems inevitably results in the mixing of residual solution within the macropores with the new solution entering the flow channels (Figure 1b). Such mixing compromises the energy efficiency of ES and, in the case of selective ES, also undermines the separation selectivity[23]. In this work, we focus primarily on the adverse impacts of mixing on the performance of ES-based desalination (i.e., we do not investigate its impact on ES-based selective separation).

The flow channels of the ES cell are alternately occupied by the diluate stream in the charging half-cycle and by the brine stream in the discharge half-cycle. There are two instances of mixing due to solution-switch in a complete ES cycle for desalination. At the end of the charging half-cycle, the salt concentration of solution in the electrode macropores equals that of the diluate (Figure 1, b1). Upon introduction of the receiving solution (at feed salinity), mixing occurs between the residual solution in the macropores with a low concentration and the receiving solution with a higher concentration (Figure 1, b2). At the end of discharge half-cycle, the salt concentration in the macopores equals that of the brine (Figure 1, b3). The brine remaining in the electrodes macropores mixes with the feed solution entering the flow channels, raising the concentration of the solution in the feed channel beyond the feed salinity at the start of the charging half-cycle (Figure 1, b4).

In this study, we perform numerical modeling to systematically evaluate the impacts of solution-switch on the performance of ES-based desalination. Specifically, we quantify the extent to which mixing due to solution-switch compromises the ability of ES to produce freshwater under both single-pass and semi-batch operation modes. We also investigate how the impacts of solution-switch on salt removal and energy consumption depend on several key parameters, including feed salinity, water recovery, electrode macroporosity, loading, and capacity.

## Theory

### Modified-Donnan theory for the salt adsorption in micropores

We apply the modified-Donnan model (mD) to simulate salt adsorption and desorption in porous AC electrodes[2,26]. The ions are mainly stored in the electrical double layers (EDLs) in the micropores of the porous AC electrode. The solution in the macropore has the same composition as the bulk solution and charge neutrality is maintained for both macropores and the bulk solution. For 1:1 salt such as NaCl, the ion concentration in the micropores, $c_{mi,i}$, relates to the salt concentration in the macropores, $c_{mA}$, according to

$$c_{mi,i} = c_{mA} \exp(-z_i \Delta\phi_D) \tag{1}$$

where $z_i$ is the ion valence and $\Delta\phi_D$ is the Donnan potential between the micropores and macropores. The subscript "$i$" is ignored for $c_{mA}$ for symmetric salt due to charge neutrality in the macropores. For symmetric salt, the charge density, defined as the net charge in the micropores ($\sigma_{mi} = \sum_i z_i c_{mi,i}$, $z_i$ is the valence of species "$i$"), can be expressed as

$$\sigma_{mi} = -2c_{mA}\sinh(\Delta\phi_D) \tag{2}$$

The total ion concentration in the micropores, $c_{mi,tot}$ ($c_{mi,ions} = \sum_i c_{mi,i}$), can be obtained by combining equations 1 and 2:

$$c_{mi,tot}^2 = \sigma_{mi}^2 + (2c_{mA})^2 \tag{3}$$

The micropore charge density, $\sigma_{mi}$, can also relate to the Stern layer potential drop, $\Delta\phi_s$, according to the following equation:

$$\sigma_{mi} F = -C_{st,vol} \Delta\phi_s V_T \tag{4}$$

where $F$ is the is the Faraday constant, $C_{st,vol}$ is the volumetric Stern layer capacitance (F L$^{-1}$), and $V_T$ is the thermal voltage (25.7 mV at 25 °C). $C_{st,vol}$ can be estimated using an empirical expression reported in literature:

$$C_{st,vol} = C_{st,vol,0} + \alpha \sigma_{mi}^2 \qquad (5)$$

where $C_{st,vol,0}$ is the capacitance in the zero-charge limit and $\alpha$ is an empirical coefficient for the charge dependence of Stern capacitance.

Assuming the electrode is saturated at the end of the charging half-cycle, there is no transport of ions across the spacer channel and the electrode. At equilibrium, the sum of $\Delta\phi_D$ and $\Delta\phi_s$ equals half of the potential across the cell for symmetric cells:

$$\Delta\phi_D + \Delta\phi_s = \frac{V_{cell}}{2} \qquad (6)$$

Combining equations 1 to 6, we can obtain the equilibrium charge density and salt concentration in the macropore under a given applied voltage. ES can be operated using many different schemes. For simplicity, we choose to analyze an ES process where charging is performed using a constant voltage until saturation and discharge is performed via short circuiting (i.e., zero voltage discharge).

### Simplified 1-D dynamic ES transport model

To obtain the concentration profile in the diluate channel as a function of time, we employ simplified 1-D dynamic ES transport model[2,27]. The ion flux, $J_i$, through the macropores is given by the Nernst-Planck equation:

$$J_i = -D_i \left( \frac{\partial c_{mA}}{\partial x} + z_i c_{mA,i} \frac{\partial \phi_{mA}}{\partial x} \right) \qquad (7)$$

where $D_i$ is the diffusion coefficient, $\phi_{mA}$ is the macropore potential, and $x$ is the coordinate starting from the spacer-electrode boundary to the electrode-current collector boundary. Across the electrode, the mass balance of ion "$i$" can be written as:

$$\frac{\partial}{\partial t}(p_{mA} c_{mA} + p_{mi} c_{mi,i}) = -p_{mA} \frac{\partial J_i}{\partial x} \qquad (8)$$

where $p_{mA}$ and $p_{mi}$ are the macropore and micropore porosity, and $c_{mi,i}$ is the concentration of species $i$ in micropores. The electrode potential is a constant, yielding:

$$\frac{\partial(\phi_{mA} + \phi_D + \phi_s)}{\partial x} = 0 \qquad (9)$$

As ion transport leads to desalination in the spacer channel, the mass balance of salt can be written as

$$p_{sp} \frac{\partial c_{sp}}{\partial t} = -\frac{J_{ions}}{L_{sp}} + \frac{c_{sp,in} - c_{sp}}{\tau_{sp}} \quad (10)$$

where $p_{sp}$ is the spacer channel porosity, $c_{sp}$ is the spacer channel salt concentration, $L_{sp}$ is the thickness of spacer channel, and $\tau_{sp}$ is the hydraulic retention time in the spacer channel. $c_{sp,in}$ is the spacer channel influent concentration. In the single-pass mode, $c_{sp,in}$ is a constant and equals the feed concentration. In the semi-batch mode with recirculation, $c_{sp,in}$ is the concentration in the diluate or concentrate tank and can be expressed as:

$$\frac{\partial c_{sp,in}}{\partial t} = \frac{c_{sp} - c_{sp,in}}{\tau_{tank}} \quad (11)$$

where $\tau_{tank}$ is the hydraulic retention time in the tank. Lastly, we also consider ion flux and current continuities at the spacer-electrode interface:

$$J_{ions} = \sum_i J_i(x = 0) \quad (12)$$

$$\frac{I}{F} = \sum_i z_i J_i(x = 0) \quad (13)$$

Solving Eqs. 1-13 yields the feed tank concentration, micropore charge density, and macropore salt concentration as a function of time. The specific energy consumption ($SEC$) of an ES process can be calculated using the following equation for short-circuit discharge where no energy is recovered:

$$SEC = \frac{1}{v_{tot}} \sum_{k=1}^{m} \int_{t_{k,0}}^{t_{k,hc}} V_{cell}(t) I(t) dt \quad (14)$$

where $v_{tot}$ is the total volume of the dilute and concentrate water, $k$ is the cycle number, $t_{k,0}$ and $t_{k,hc}$ is the initial and final time point of the $k$th charging half-cycle, $V_{cell}(t)$ and $I(t)$ are cell voltage and current at time $t$. Constant voltage in the charging stage is employed in this study. The final salt removal ($SR$) is obtained as:

$$SR = 1 - \frac{C_d}{C_f} \quad (15)$$

where $C_f$ and $C_d$ are the concentrations of the feed and diluted stream. $SR$ is equivalent to the salt rejection in reverse osmosis or nanofiltration. The parameters utilized in the calculation are selected from ranges commonly reported or simulated in literature, as listed in Table 1. These parameters in Table 1 are used throughout all simulations except when the parameters are the subject of investigation.

**Table 1.** Parameters for simulation

| Symbols | Description | Value | Dimension | Reference |
|---|---|---|---|---|
| $C_{st,vol,0}$ | Zero-charge capacitance | 200 | F g$^{-1}$ | 28 |
| $\alpha$ | Charge dependence of Stern capacitance | 20 | | 27 |
| $V_{cell}$ | Cell voltage | 1.2 | V | 1 |
| $\rho_{elec}$ | Electrode mass density | 0.75 | g ml$^{-1}$ | 29 |
| $p_{sp}$ | Spacer porosity | 0.5 | | 17 |
| $p_{mi}$ | Microporosity | 0.25 | | 30 |
| $p_{sp}$ | Macroporosity | 0.5 | | 31 |
| $L_{elec}$ | Electrode thickness | 250 | μm | 32 |
| $L_{sp}$ | Spacer thickness | 250 | μm | 33 |
| $T_{hc}$ | Half-cycle time | 30 | min | 34 |

**Mass balance in ES with or without mixing**

In the dynamic ES transport model, the effect of mixing is studied by setting the initial boundary conditions of macropore concentration for the charging and discharge half-cycles. Without mixing, the concentration in the macropores at beginning of charging or discharge half-cycle equals the concentration of the diluate or the concentrate stream.

With mixing, the concentration in the macropores at the beginning of the charging half-cycle equals the concentration of the brine, as the pores preserve the brine from the previous discharge half-cycle. Similarly, the concentration in the macropores at the beginning of the discharge half-cycle equals the concentration of the diluate.

In addition, analytical equations can be derived to relate the salt concentration in diluate, brine, the macropores, and micropores at the beginning of charging or discharge half-cycle and these concentrations at equilibrium with mass-balance equations. These analytical equations (Eqs. 16–19) are used to investigate the influence of mixing under different operational conditions and electrode properties on salt removal in a single-cycle batch-mode due to their simplicity. Without mixing, the concentration in the macropores at beginning of charging or discharge half-cycle equals the concentration of the feed. The mass balance holds for both charging (Eq. 16) and discharge half-cycle (Eq. 17):

$$nc_f(v_d + v_{mA}) + c_b v_{mi} = c_{mi,ions} v_{mi} + nc_{mA}(v_{mA} + v_d) \tag{16}$$

$$c_{mi,ions} v_{mi} + nc_f(v_{mA} + v_b) = nc_b(v_b + v_{mA} + v_{mi}) \tag{17}$$

where $n$ is the van't Hoff factor (e.g., $n$ equals 2 for NaCl), $c_0$ is the feed concentration, $v_{di}$ is the volume of the diluate (in the flow channel, tubing, and the tank), and $v_{mA}$, and $v_{mi}$ are the volumes of electrode macropores and micropores. The product of concentration and volume at the beginning of the charging (or discharge) half-cycle (the left-hand side of Eqs. 16-19) equals the product of concentration and volume at the end of the charging (or discharge) half-cycle (the right-hand side of Eqs. 16-19). With mixing, we apply mass balance for the charging (Eq. 18) and discharge (Eq. 19) half-cycle:

$$nc_f v_d + nc_b(v_{mA} + v_{mi}) = c_{mi,ions} v_{mi} + nc_{mA}(v_{mA} + v_d) \tag{18}$$

$$nc_f v_b + c_{mi,tot} v_{mi} + nc_{mA} v_{mA} = nc_b(v_b + v_{mA} + v_{mi}) \tag{19}$$

where $c_b$ is the brine concentration and $v_b$ is the volume of brine tank. The volume of the macropores ($v_{ma}$) and micropores ($v_{mi}$) are given by the

$$v_{mi} = p_{mi} v_{elec} \tag{20}$$

$$v_{mA} = p_{mA} v_{elec} \quad (21)$$

$$v_{elec} = \frac{m_{elec}}{\rho_{elec}} \quad (22)$$

where $p_{mi}$ and $p_{mA}$ are the microporosity and macroporosity, and $v_{elec}, m_{elec}, \rho_{elec},$ are the electrode volume, mass, and density. Combining Eqs. 1–6 and 15–22, we can obtain the salt removal with or without mixing at a given operational conditions without solving the temporal and spatial evolution of concentration in the charging and discharging half-cycle in single-cycle semi-batch mode.

**Single-pass vs semi-batch operation modes in ES**

Based on whether the solutions are recirculated between their respective tanks and the ES cell, we can categorize film electrode-based ES into two modes: semi-batch and single-pass modes (Figure 2). In the semi-batch mode, the diluate stream is recirculated between the ES cell and the diluate tank, whereas the brine stream is recirculated between the ES cell and the brine tank. Semi-batch ES can have one or multiple charging-discharge cycles depending on the desalination goal and the operational parameters. Semi-batch ES with a single flow cycle represents a process where the diluate stream is recirculated until the end of a charging cycle and then collected as the product water (Figure 2a). This is practical only if a small salinity reduction needs to be achieved or/and a large electrode mass with high salt adsorption capacity is used to treat a small volume of feed solution. Alternatively, a semi-batch ES can also be operated using multiple charging-discharge cycles where the diluate desalinated in previous charging cycles continues to be desalinated in subsequent charging cycles after the electrodes are regenerated in the discharge cycles. Only when the salinity of the diluate reaches a sufficiently low target level will the diluate be collected as the product water (Figure 2a).

In the single-pass mode, the feed solution flows through the ES cell once and flows out as the diluate as the ions are removed from the solution and stored in the electrode temporarily. When the electrode is charged to the desired level (e.g., saturated at an applied voltage, or voltage becoming too high at an applied current), the cell is either short-circuited or a reverse voltage is applied to release the stored ions to the feed solution, resulting in a brine stream (Figure 2 b). The key feature of the single-pass mode is that the effluents in charging or discharge half-cycles are not sent back to the tanks from which their respective influents are drawn from. For an ES process with a single-pass mode to generate a significant salinity reduction (from the feed solution to diluate), it requires a long hydraulic residence time and a relatively large electrode mass with high salt adsorption capacity for a given volume of solution desalinated.

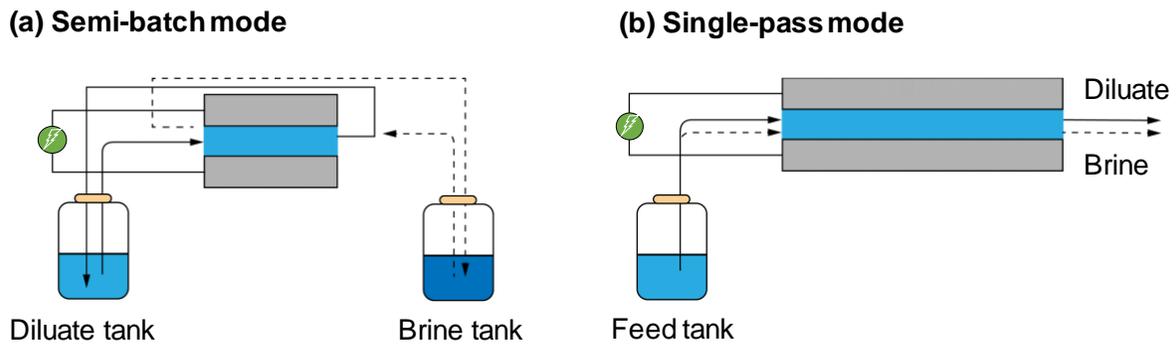

| Feature | Semi-batch | | Single-pass |
| --- | --- | --- | --- |
| | Single-cycle | Multi-cycle | |
| Recirculation | Yes | Yes | No |
| Electrode to salt ratio | Large | Small | Large |
| Number of half-cycles* before diluate (or brine) is collected as product | One | Multiple | One |
| Mixing Occurrence | Batches | Half-cycles* and batches | Half-cycles* |
| Influent concentration | Time-dependent | Time-dependent | Constant: $C_f$ |

Note: *Half-cycles refers to charging and discharge half-cycles.

**Figure 2.** Schematics and feature comparison of (a) semi-batch mode and (b) single-pass mode in ES. The semi-batch mode can be further divided into a single-cycle semi-batch mode and a multi-cycle semi-batch mode, depending on the number of charging/discharge cycles a batch of solution is treated for before it is collected as the product freshwater.

## Results and Discussion

**Impact of mixing on desalination performance in single-cycle semi-batch mode**

In the single-cycle semi-batch mode, a large electrode mass (or area) is necessary to reach target salinity reduction, which in turn retains a large volume of solution in the macropores having the effluent salinity at the end of the preceding half-cycle. At the beginning of the charging half-cycle, the feed solution entering the flow channel will experience a concentration jump due to the mixing between the feed solution and the brine remaining in the macropores at the end of the discharge half-cycle. Similarly, at the beginning of the discharge half-cycle, mixing occurs between the influent with the feed salinity and the diluate remaining in the macropores at the end of the preceding charging half-cycle, which results in a sudden concentration drop in the brine flow. These effects are evident when comparing salinity of the solution in the channel at the beginning of the charging and discharge half-cycles to the influent salinity (Figure 3a). With the parameters used in this investigation, the presence of mixing reduces the salt removal from 40% to 32% for a 100 mM NaCl solution (typical brackish water salinity).

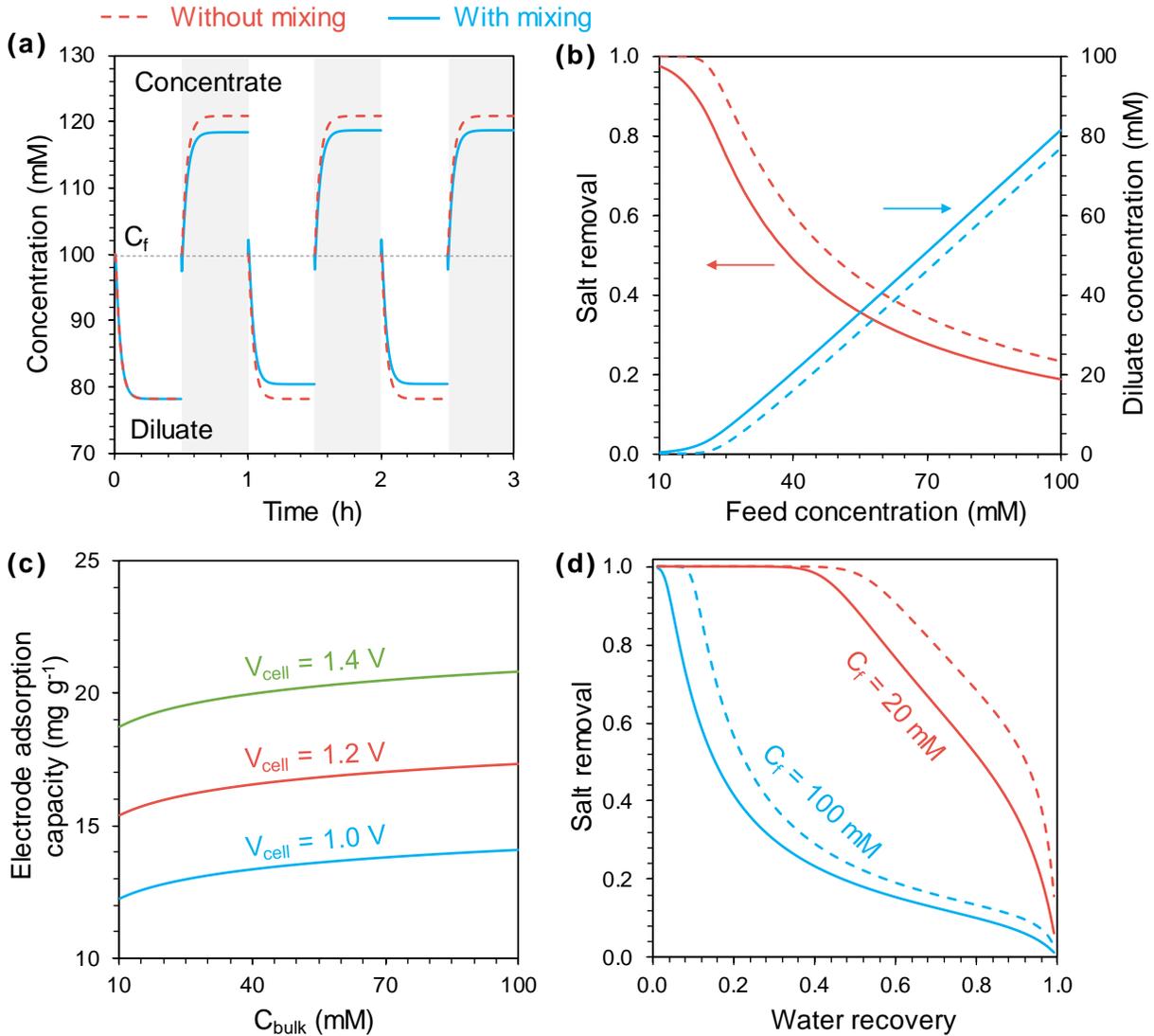

**Figure 3.** (a) Concentration of diluate and brine as a function of time in an ES process with single-cycle semi-batch mode without mixing (red dashed curve) or with mixing (blue solid curve). (b) Influence of mixing on ES performance with varying feed concentration in single-cycle semi-batch mode. (c) Dependence of electrode adsorption capacity on cell voltage and bulk salinity. (d) Influence of mixing on ES performance with varying water recovery in single-cycle semi-batch mode. The results are simulated using the following parameters: the feed concentration is 100 mM (a), and the electrode loadings, defined as electrode mass per feed volume, is 100 g L$^{-1}$ (a, c, d). Other parameter settings can be found in table 1.

The impact of mixing on ES performance varies with operational conditions and electrode properties. With other parameters kept constant, the salt removal decreases with increasing feed concentration. An ES system that can achieve near complete desalination of a low-salinity feed solution (e.g., 10 mM) can only achieve a salt removal of ~20% when the feed salinity increases to 100 mM (Figure 3b). Compared to the ideal scenario without mixing, the salt removal is systematically undermined by mixing due to solution-switch. The salt removal with mixing is around 80% of that without mixing with a feed concentration of 30-100 mM. At adsorption equilibrium, the electrode adsorption capacity increases with increasing bulk solution concentration, and increasing the applied voltage also improves the electrode adsorption capacity (Figure 3c). However, the electrode capacity is only tens of milligram of salt per gram of AC electrode under commonly used operation parameters,[35] which explains the relatively low salt removal for a feed even with a moderate salinity (e.g., 100 mM, as shown in Figure 3b).

The analysis shown in Figure 3b assumes a 50% water recovery ($WR$), defined as the ratio between the diluate volume and the total volume of diluate and brine. With a low feed concentration (<20 mM), the salt removal first remains high (near 100%) with a $WR$ below 40% as the electrodes have not reached saturation. Salt removal declines sharply when further increasing $WR$ due to electrode saturation. With a feed concentration of 100 mM, salt removal decreases with increasing $WR$ (Figure 3d). With increasing $WR$, the volume of the diluate and thus the amount of salt to be removed increase but the capacity of the electrode is fixed at a certain voltage. At the same time, these ions need to be concentrated in a smaller volume of brine and thus generates a more concentrated brine. This explains the tradeoff between salt removal and $WR$. The presence of mixing further compromises salt removal by over 20% when $WR$ exceeds 7% with a feed concentration of 100 mM.

Without mixing, salt removal is unaffected by macroporosity. With mixing, however, salt removal decreases with increasing macroporosity (Figure 4a), because a larger volume (relative to the spacer channel volume) of solution remaining in the macropores will mix with the influent solution. The percentage reduction of salt removal due to mixing also increases with increasing microporosity. Despite this negative impact, the presence of macropores is necessary because they provide fast transport pathways for ions to access the micropores.[36] Decreasing macroporosity will thus increase ion transport resistance by reducing the "highways" for ions to transport from the spacer channel to the micropores.

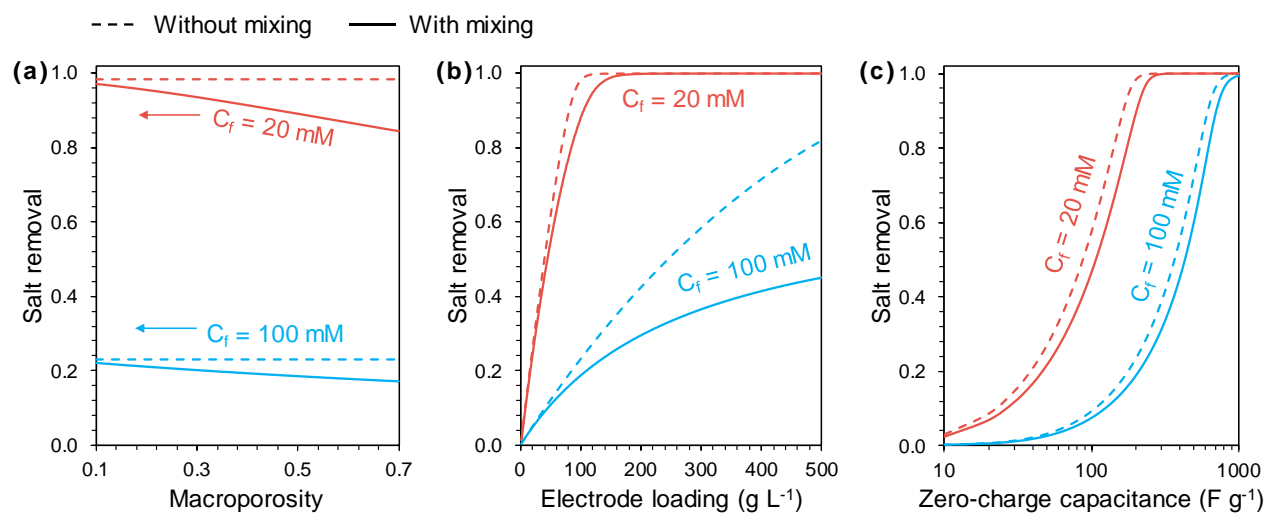

**Figure 4.** Influence of mixing on ES salt removal with varying macroporosity (a), electrode loading (b), and electrode Stern capacitance at zero-charge limit (c) in single-cycle semi-batch mode. The impact was simulated under two feed concentrations: 20 mM and 100 mM. The electrode loadings used in the calculation is 100 g L$^{-1}$. Other parameter settings can be found in table 1.

To enhance salt removal, one effective strategy is to increase the electrode mass (or area) relative to the treated water volume. This strategy is generally feasible without considering the influence of mixing (Figure 4b). The salt removal of a 100 mM feed solution can be elevated to above 80%, though with an impractically large electrode to solution ratio (e.g., 500 g electrode per liter of feed solution). However, in the presence of mixing during solution-switch, even the benefit of using more electrode is compromised due to the increased macropore volume. In other words, although a large electrode mass can store more salts per charging half-cycle, it also carries in the macropores a larger volume of solution that later becomes available for mixing during solution-switch, and thus cannot mitigate the negative impact of mixing. With 500 g of electrode for treating 1 L of feed solution (100 mM), the salt removal is only 45% when the effect of mixing is considered, vs. 82% in the ideal scenario without mixing.

Increasing electrode capacity effectively improves salt removal but cannot eliminate the influence of mixing (Figure 4c). Carbon electrodes reported in the literature have zero-charge capacitances ranging from 50 to 500 F $g^{-1}$,[37] which can only desalinate low-salinity brackish water (e.g., 20–65 mM) but fail to produce freshwater from brackish water with higher salinity. Although much effort has been devoted to developing high-capacity electrodes in the past decade, to achieve a salt removal over 90% from 1L of a 100 mM solution (NaCl) using 100 g of electrodes requires the electrode capacity to exceed 700 F $g^{-1}$. Using ES for seawater desalination is even more challenging and will require electrodes with capacity that is at least an order of magnitude higher than that of the state-of-the-art.

**Impact of mixing on desalination performance in multi-cycle semi-batch mode**

As shown in the above section, a unit mass of electrode can only store a small number of ions during a charging half-cycle, which means an extremely large mass of electrode (per volume of solution) is needed to reach the target salt removal to desalinate brackish water even with a medium concentration. To address this challenge, multi-cycle semi-batch mode has been employed in ES where the same feed solution (or brine) flows through the ES cell for multiple charging (or discharge) half-cycles, with salt removed (or released) incrementally each cycle to achieve a significant cumulative salt removal.[21,38]

In the multi-cycle semi-batch mode, a small electrode mass can be used, which results in a small macropore volume as compared to the solution volume. However, the smaller relative macropore volume does not alleviate the detrimental impact of the mixing due to solution-switch because it takes more cycles to achieve the target desalination and the effect of mixing will build up over multiple cycles (Figure 5a). Without mixing, we can increase the number of cycles number to achieve the desired level of salinity reduction. With mixing, however, the salt removal in each cycle decreases due to the increasingly severe mixing as the concentration difference between the brine and diluate grows. For example, the salt removal of a 100 mM NaCl solution after 5 cycles reaches 99% without mixing whereas only 65% of salt is removed with mixing (Figure 5a). Mixing becomes increasingly severe as we achieve a more complete desalination (i.e., a larger difference between the diluate and brine concentrations), as evidenced by the growingly larger "spikes" observed at the solution-switch steps.

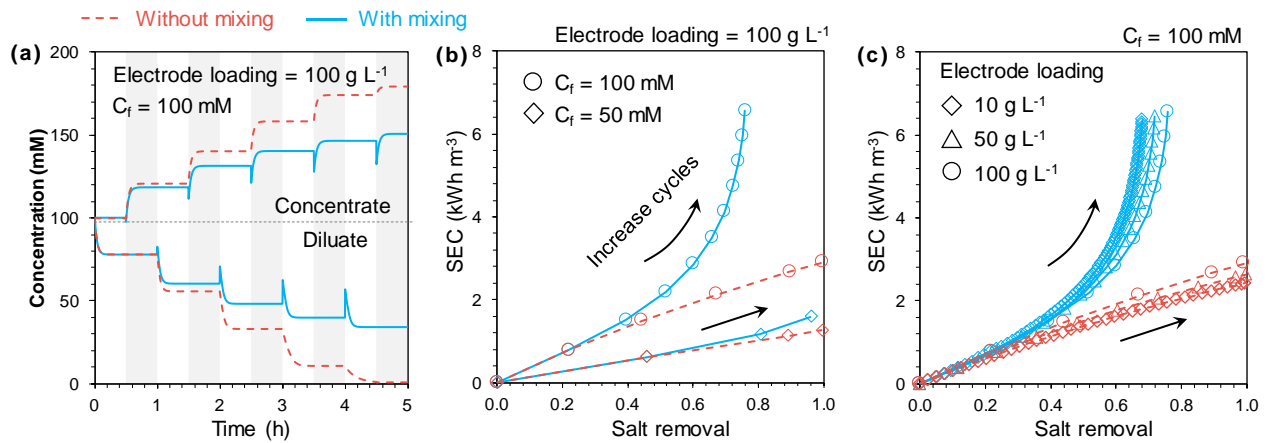

**Figure 5.** (a) Concentration of diluate and brine as a function of time in multi-cycle semi-batch mode without mixing (red dashed curve) or with mixing (blue solid curve). The spikes in the blue curves represent the degree of mixing due to solution switch. (b, c)The relationship of specific energy consumption ($SEC$) and salt removal in multi-cycle semi-batch mode without mixing (red dashed curve) and with mixing (blue solid curve) for (b) different feed concentrations (50 and 100 mM) and (c) different electrode loading (10, 50, and 100 g L$^{-1}$). In both (b) and (c), the increase in salt removal is achieved by increasing the charging/discharge cycles. The key parameters for simulation are shown in the figures whereas the remaining parameters use values presented in Table 1.

The influence of mixing on the salt removal and specific energy consumption ($SEC$) is further evaluated in multi-cycle semi-batch mode. The salt can be fully removed from feed in the absence of mixing with increasing the number of charging-discharge cycles

(red curves in Figures 5b, 5c). For example. with a constant electrode loading of 100 g L$^{-1}$, the removal for a feed solution of 50 mM and 100 mM NaCl increase from 46% and 22% with a single cycle to over 99% after 3 cycles and 5 cycles without mixing (Figures 5b). However, with mixing, the removal is only 96% and 65% after 3 cycles and 5 cycles for the same feed solutions (Figures 5b).

The presence of mixing not only slows down the desalination rate, but also limits the salt removal and increases $SEC$. Further increasing the cycle number from 5 to 10 only marginally increase the removal for a 100 mM NaCl solution from 65% to 75% though the $SEC$ almost doubles (Figures 5b). $SEC$ without mixing is roughly linear to salt removal (Figures 5b, 5c), but $SEC$ with mixing increases much more rapidly as salt removal increases, especially with a higher feed concentration. Though close to $SEC$ and salt removal without mixing in the initial cycles, $SEC$ and salt removal with mixing increasingly deviates from the ideal case as cycle number increases. The presence of mixing during solution-switch compromises the benefit of increasing cycle number in multi-cycle semi-batch mode, which is anaglous to the reduced benefit of increasing electrode loading in single-cycle semi-batch mode.

The $SEC$ is similar for different electrode loadings to achieve the same salt removal (Figures 5c), again corroborating our previous observation that increasing electrode loading cannot effectively mitigate the detrimental impact of mixing. Like the single-cycle semi-batch mode, the benefit of elevating the salt removal by using a higher electrode loading in multi-cycle semi-batch mode is also compromised due to increased macropore volume available for mixing. In summary, even though multi-cycle semi-batch mode can achieve the same target level of desalination with less electrode loading as compared to the single-cycle semi-batch mode, the impacts of electrode and system properties on the ES desalination performance due to the mixing effect are similar in both modes.

**Impact of mixing on salt removal in single-pass mode**

In most laboratory-scale single-pass experiments performed using constant voltage charging and short-circuit discharge, the detrimental impact of solution-switch can be rightfully ignored because there is minimum mixing with such an operation. Specifically, in most of such experiments, charging is performed until the electrodes are saturated (e.g., adsorption equilibrium is reached) and the flow channel is flushed with feed solution without any salt removed. Therefore, there is no salinity difference between the solutions flowing through the ES cell before and after solution-switch. In other words, the flow channel is effectively flushed by the feed solution toward the end of both the charging and discharge half-cycles when neither charging nor discharge is actively performed (Figures 6a).

In such a context, solution-switch does not induce mixing, yet flushing itself reduces the salt removal in the charging half-cycle and results in more wasted water in the discharge half-cycle. For example, the cumulative salt removal of a 30 min half-cycle time is only 24%; in comparison, the cumulative salt removal is 66% without mixing with a half-cycle time of 5 min in single-pass mode. Without extensive flushing, the impact of mixing between the remaining solution in macropores and the influent feed solution cannot be ignored. For example, mixing reduces the maximum salt removal for a feed solution of 100 mM NaCl from 66% to 56% with a half-cycle time of 5 min (Figures 6b).

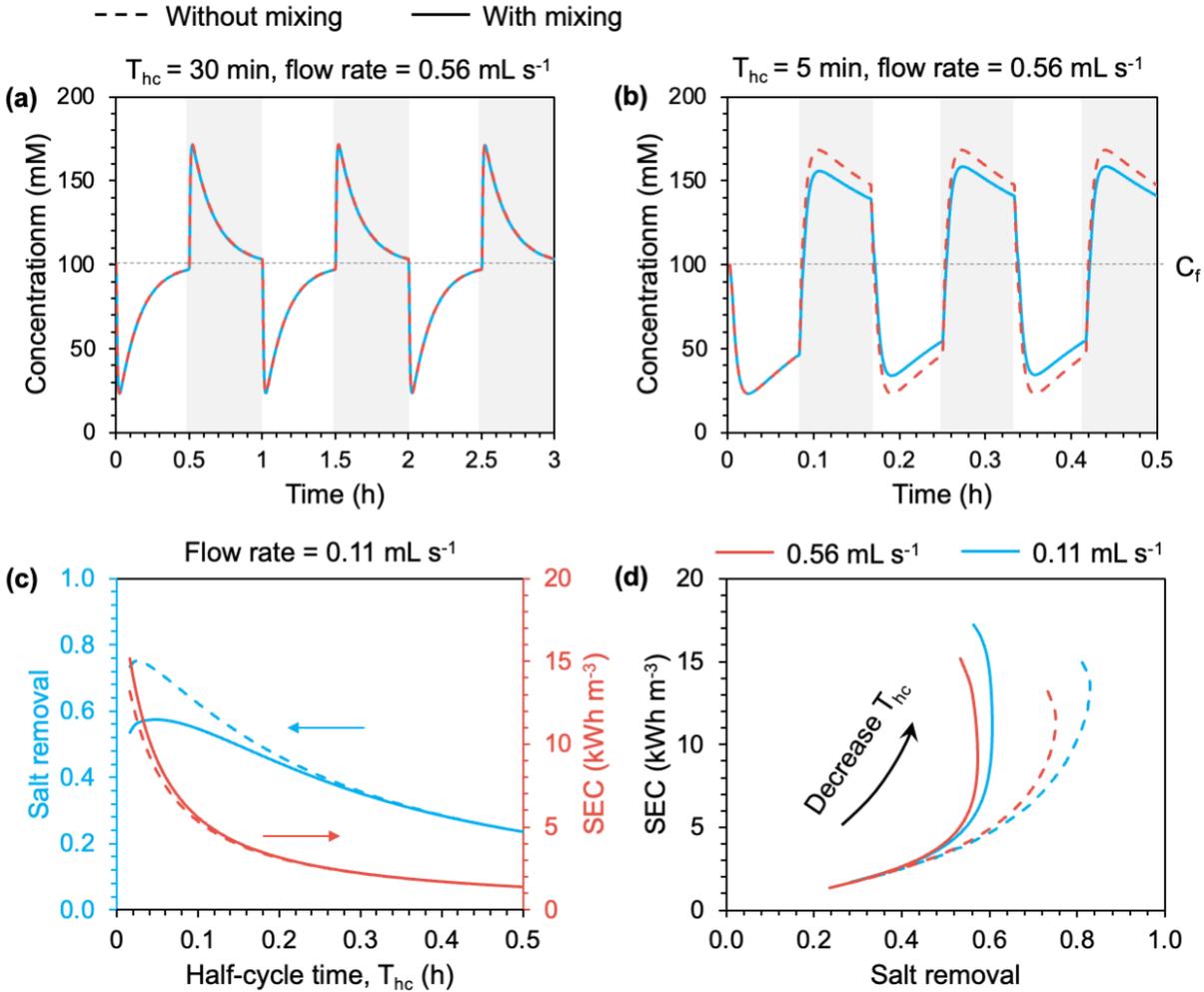

**Figure 6.** Concentrations of diluate and brine stream effluents as a function of time in single-pass mode without mixing (dashed curve) or with mixing (solid curve) with a half-cycle time ($T_{hc}$) of (a) 30 min and (b) 5 min. (c) Dependence of salt removal and $SEC$ on half-cycle time without mixing (dashed curve) or with mixing (solid curve). (d) Relationship between $SEC$ and salt removal as half-cycle time varies, for two different flow rates, with and without mixing. The results are simulated all using 100 mM NaCl as the feed solution. Other parameter settings can be found in Table 1.

For practical applications of ES with the goal of producing fresh water, the charging should be operated at optimum half-cycle time to obtain high removal with low $SEC$. Both salt removal and $SEC$ increase with decreasing half-cycle time with a fixed electrode mass and flow rate (Figures 6c). The adverse effect of mixing becomes more severe with increasing salt removal as half-cycle time decreases and a higher salt removal is achieved, but the impact of mixing on salt removal is much more significant than that on $SEC$. Increasing the flowrate with the same electrode mass does not mitigate the impact of mixing on salt removal when a relatively high salt removal is achieved with a very short half-cycle time (Figures 6d). In other words, in a practical single-pass ES system with a realistic electrode mass per volume of treated water, there is no viable means of achieving a high salt removal if the feed salinity is moderately high. If we use a long half-cycle time, the effect of mixing is mitigated by extensive flushing, but the salt removal is low due to the long half-cycle; if we use a short half-cycle which supposedly yields a high salt removal in the ideal scenario without mixing, mixing then plays an important role which reduces the salt removal. We note that the ES with constant current operation will experience similar mixing due to solution-switch as that situation described in Figure 6b and the short half-cycle time regime in Figure 6d.

## Perspective

Due to the limited adsorption capacity of film electrodes, ES systems require circuit reversal and solution-switch to sustain desalination. Regardless of single-cycle semi-batch mode, multi-cycle semi-batch mode, or single-pass mode (without extensive flushing), mixing between the influent solution and solution remaining in the electrode macropores will lead to appreciable performance deterioration. The detrimental impacts on performance due to mixing are manifested by reduced salt removal and/or increased specific energy consumption. Such detrimental impacts are more severe when the diluate and brine concentrations are more different, which occurs when the feed salinity is high, the salinity reduction is significant, and/or the water recovery is large. Merely increasing the electrode loading per volume of solution treated is ineffective in mitigating the detrimental impacts of mixing. Increasing the specific adsorption capacity of electrodes helps alleviating the detrimental impacts of mixing. While research on ES has been extensively devoted to developing electrodes with higher capacity,[39,40] the analysis presented in this work is likely the first in elucidating the true relevance of capacity to desalination performance. After all, the goal of ES is separation, not storage of ions.

In addition to desalination, research efforts have been growing to unlock new opportunities for ES in other applications such as mineral extraction and resource recovery. For example, ES based on Li-intercalation electrode materials has been shown to be capable of highly selective separation with feed solutions of complex composition and high salinity, which makes ES a promising technological candidate for chemical-free direct lithium extraction. [11,12] Film electrode-based ES for Li extraction also faces the challenge of mixing due to solution-switch, which has additional detrimental impacts on selectivity because macropores store ions non-selectively and will release unwanted ions to the receiving solution.[23] A common approach to minimize the adverse effects of mixing on selectivity is to add a rinsing step before solution-switch to remove solutions stored in the macropores. Such a rinsing step can improve the purity of lithium in the receiving solution but at the cost of more freshwater consumption. Addressing the negative impacts of solution-switch in various applications of ES can benefit from developing electrodes with higher capacity, but only to a limited extent. Future research should explore new configurations and operational modes to mitigate the detrimental impacts of mixing due to solution-switch.

# References

(1) Suss, M. E.; Porada, S.; Sun, X.; Biesheuvel, P. M.; Yoon, J.; Presser, V. Water Desalination via Capacitive Deionization: What Is It and What Can We Expect from It? *Energy Environ Sci* **2015**, *8* (8), 2296–2319. https://doi.org/10.1039/c5ee00519a.

(2) Porada, S.; Zhao, R.; Van Der Wal, A.; Presser, V.; Biesheuvel, P. M. Review on the Science and Technology of Water Desalination by Capacitive Deionization. *Prog Mater Sci* **2013**, *58* (8), 1388–1442. https://doi.org/10.1016/J.PMATSCI.2013.03.005.

(3) Zhao, R.; Biesheuvel, P. M.; Van Der Wal, A. Energy Consumption and Constant Current Operation in Membrane Capacitive Deionization. *Energy Environ Sci* **2012**, *5* (11), 9520–9527. https://doi.org/10.1039/c2ee21737f.

(4) Wang, L.; Dykstra, J. E.; Lin, S. Energy Efficiency of Capacitive Deionization. *Environ Sci Technol* **2019**, *53* (7), 3366–3378. https://doi.org/10.1021/acs.est.8b04858.

(5) Tsai, S. W.; Hackl, L.; Kumar, A.; Hou, C. H. Exploring the Electrosorption Selectivity of Nitrate over Chloride in Capacitive Deionization (CDI) and Membrane Capacitive Deionization (MCDI). *Desalination* **2021**, *497*, 114764. https://doi.org/10.1016/J.DESAL.2020.114764.

(6) Xu, L.; Yu, C.; Tian, S.; Mao, Y.; Zong, Y.; Zhang, X.; Zhang, B.; Zhang, C.; Wu, D. Selective Recovery of Phosphorus from Synthetic Urine Using Flow-Electrode Capacitive Deionization (FCDI)-Based Technology. *ACS Environmental Science and Technology Water* **2021**, *1* (1). https://doi.org/10.1021/acsestwater.0c00065.

(7) Lin, L.; Hu, J.; Liu, J.; He, X.; Li, B.; Li, X. Y. Selective Ammonium Removal from Synthetic Wastewater by Flow-Electrode Capacitive Deionization Using a Novel K2Ti2O5-Activated Carbon Mixture Electrode. *Environ Sci Technol* **2020**, *54* (19). https://doi.org/10.1021/acs.est.0c04383.

(8) Gamaethiralalage, J. G.; Singh, K.; Sahin, S.; Yoon, J.; Elimelech, M.; Suss, M. E.; Liang, P.; Biesheuvel, P. M.; Zornitta, R. L.; De Smet, L. C. P. M. Recent Advances in Ion Selectivity with Capacitive Deionization. *Energy and Environmental Science*. 2021. https://doi.org/10.1039/d0ee03145c.

(9) Porada, S.; Zhao, R.; Van Der Wal, A.; Presser, V.; Biesheuvel, P. M. Review on the Science and Technology of Water Desalination by Capacitive Deionization. *Prog Mater Sci* **2013**, *58* (8), 1388–1442. https://doi.org/10.1016/J.PMATSCI.2013.03.005.

(10) Xing, W.; Liang, J.; Tang, W.; He, D.; Yan, M.; Wang, X.; Luo, Y.; Tang, N.; Huang, M. Versatile Applications of Capacitive Deionization (CDI)-Based Technologies. *Desalination* **2020**, *482*, 114390. https://doi.org/10.1016/J.DESAL.2020.114390.

(11) Liu, C.; Li, Y.; Lin, D.; Hsu, P. C.; Liu, B.; Yan, G.; Wu, T.; Cui, Y.; Chu, S. Lithium Extraction from Seawater through Pulsed Electrochemical Intercalation. *Joule* **2020**, *4* (7). https://doi.org/10.1016/j.joule.2020.05.017.

(12) Battistel, A.; Palagonia, M. S.; Brogioli, D.; La Mantia, F.; Trócoli, R. Electrochemical Methods for Lithium Recovery: A Comprehensive and Critical Review. *Advanced Materials*. 2020. https://doi.org/10.1002/adma.201905440.

(13) Zhang, C.; He, D.; Ma, J.; Tang, W.; Waite, T. D. Comparison of Faradaic Reactions in Flow-through and Flow-by Capacitive Deionization (CDI) Systems. *Electrochim Acta* **2019**, *299*. https://doi.org/10.1016/j.electacta.2019.01.058.

(14) Guyes, E. N.; Shocron, A. N.; Simanovski, A.; Biesheuvel, P. M.; Suss, M. E. A One-Dimensional Model for Water Desalination by Flow-through Electrode Capacitive Deionization. *Desalination* **2017**, *415*. https://doi.org/10.1016/j.desal.2017.03.013.

(15) Gao, X.; Omosebi, A.; Landon, J.; Liu, K. Surface Charge Enhanced Carbon Electrodes for Stable and Efficient Capacitive Deionization Using Inverted Adsorption-Desorption Behavior. *Energy Environ Sci* **2015**, *8* (3). https://doi.org/10.1039/c4ee03172e.

(16) Biesheuvel, P. M.; van der Wal, A. Membrane Capacitive Deionization. *J Memb Sci* **2010**, *346* (2), 256–262. https://doi.org/10.1016/J.MEMSCI.2009.09.043.

(17) Wang, L.; Lin, S. Mechanism of Selective Ion Removal in Membrane Capacitive Deionization for Water Softening. *Environ Sci Technol* **2019**, *53* (10). https://doi.org/10.1021/acs.est.9b00655.

(18) Lee, J.; Jo, K.; Lee, J.; Hong, S. P.; Kim, S.; Yoon, J. Rocking-Chair Capacitive Deionization for Continuous Brackish Water Desalination. *ACS Sustain Chem Eng* **2018**, *6* (8). https://doi.org/10.1021/acssuschemeng.8b02123.

(19) Lee, J.; Kim, S.; Kim, C.; Yoon, J. Hybrid Capacitive Deionization to Enhance the Desalination Performance of Capacitive Techniques. *Energy Environ Sci* **2014**, *7* (11). https://doi.org/10.1039/c4ee02378a.

(20) Singh, K.; Porada, S.; de Gier, H. D.; Biesheuvel, P. M.; de Smet, L. C. P. M. Timeline on the Application of Intercalation Materials in Capacitive Deionization. *Desalination* **2019**, *455*, 115–134. https://doi.org/10.1016/J.DESAL.2018.12.015.

(21) Do, V. Q.; Reale, E. R.; Loud, I. C.; Rozzi, P. G.; Tan, H.; Willis, D. A.; Smith, K. C. Embedded, Micro-Interdigitated Flow Fields in High Areal-Loading Intercalation Electrodes towards Seawater Desalination and Beyond. *Energy Environ Sci* **2023**, *16* (7). https://doi.org/10.1039/d3ee01302b.

(22) Jeon, S. Il; Park, H. R.; Yeo, J. G.; Yang, S.; Cho, C. H.; Han, M. H.; Kim, D. K. Desalination via a New Membrane Capacitive Deionization Process Utilizing Flow-Electrodes. *Energy Environ Sci* **2013**, *6* (5). https://doi.org/10.1039/c3ee24443a.

(23) Xiong, J.; He, L.; Zhao, Z. Lithium Extraction from High-Sodium Raw Brine with $Li_{0.3}FePO_4$ Electrode. *Desalination* **2022**, *535*, 115822. https://doi.org/10.1016/J.DESAL.2022.115822.

(24) Yang, T.; Zhang, H.; Guo, L.; Wang, J.; Guo, Z.; Du, Y.; Liu, J.; Zhao, Y.; Zhang, P.; Ji, Z. Y. Mesopore-Enhanced Graphene Electrodes with Modified Hydrophilicity for Ultrahigh Capacitive Deionization. *Desalination* **2023**, *567*, 116984. https://doi.org/10.1016/J.DESAL.2023.116984.

(25) Liu, T.; Serrano, J.; Elliott, J.; Yang, X.; Cathcart, W.; Wang, Z.; He, Z.; Liu, G. Exceptional Capacitive Deionization Rate and Capacity by Block Copolymer-Based Porous Carbon Fibers. *Sci Adv* **2020**, *6* (16). https://doi.org/10.1126/sciadv.aaz0906.

(26) Wang, L.; Biesheuvel, P. M.; Lin, S. Reversible Thermodynamic Cycle Analysis for Capacitive Deionization with Modified Donnan Model. *J Colloid Interface Sci* **2018**, *512*, 522–528. https://doi.org/10.1016/J.JCIS.2017.10.060.

(27) Dykstra, J. E.; Zhao, R.; Biesheuvel, P. M.; Van der Wal, A. Resistance Identification and Rational Process Design in Capacitive Deionization. *Water Res* **2016**, *88*, 358–370. https://doi.org/10.1016/J.WATRES.2015.10.006.

(28) Kim, T.; Yoon, J. Relationship between Capacitance of Activated Carbon Composite Electrodes Measured at a Low Electrolyte Concentration and Their Desalination Performance in Capacitive Deionization. *Journal of Electroanalytical Chemistry* **2013**, *704*, 169–174. https://doi.org/10.1016/J.JELECHEM.2013.07.003.

(29) Yeh, C. L.; Hsi, H. C.; Li, K. C.; Hou, C. H. Improved Performance in Capacitive Deionization of Activated Carbon Electrodes with a Tunable Mesopore and Micropore Ratio. *Desalination* **2015**, *367*, 60–68. https://doi.org/10.1016/J.DESAL.2015.03.035.

(30) Wang, L.; Lin, S. Membrane Capacitive Deionization with Constant Current vs Constant Voltage Charging: Which Is Better? *Environ Sci Technol* **2018**, *52* (7), 4051–4060. https://doi.org/10.1021/acs.est.7b06064.

(31) Dykstra, J. E.; Keesman, K. J.; Biesheuvel, P. M.; van der Wal, A. Theory of PH Changes in Water Desalination by Capacitive Deionization. *Water Res* **2017**, *119*. https://doi.org/10.1016/j.watres.2017.04.039.

(32) Wang, L.; Lin, S. Intrinsic Tradeoff between Kinetic and Energetic Efficiencies in Membrane Capacitive Deionization. *Water Res* **2018**, *129*, 394–401. https://doi.org/10.1016/j.watres.2017.11.027.

(33) Wang, L.; Lin, S. Theoretical Framework for Designing a Desalination Plant Based on Membrane Capacitive Deionization. *Water Res* **2019**, *158*, 359–369. https://doi.org/10.1016/j.watres.2019.03.076.

(34) Kim, T.; Gorski, C. A.; Logan, B. E. Low Energy Desalination Using Battery Electrode Deionization. *Environ Sci Technol Lett* **2017**, *4* (10), 444–449. https://doi.org/10.1021/acs.estlett.7b00392.

(35) Kim, T.; Yoon, J. CDI Ragone Plot as a Functional Tool to Evaluate Desalination Performance in Capacitive Deionization. *RSC Adv* **2015**, *5* (2). https://doi.org/10.1039/c4ra11257a.

(36) Liu, R.; Chen, L.; Yao, S.; Shen, Y. Pore-Scale Study of Capacitive Charging and Desalination Process in Porous Electrodes and Effects of Porous Structures. *J Mol Liq* **2021**, *332*, 115863. https://doi.org/10.1016/J.MOLLIQ.2021.115863.

(37) Liu, Y.; Nie, C.; Liu, X.; Xu, X.; Sun, Z.; Pan, L. Review on Carbon-Based Composite Materials for Capacitive Deionization. *RSC Adv* **2015**, *5* (20). https://doi.org/10.1039/c4ra14447c.

(38) Reale, E. R.; Regenwetter, L.; Agrawal, A.; Dardón, B.; Dicola, N.; Sanagala, S.; Smith, K. C. Low Porosity, High Areal-Capacity Prussian Blue Analogue Electrodes Enhance Salt Removal and Thermodynamic Efficiency in Symmetric Faradaic Deionization with Automated Fluid Control. *Water Res X* **2021**, *13*. https://doi.org/10.1016/j.wroa.2021.100116.

(39) Huang, Z. H.; Yang, Z.; Kang, F.; Inagaki, M. Carbon Electrodes for Capacitive Deionization. *Journal of Materials Chemistry A*. 2017. https://doi.org/10.1039/c6ta06733f.

(40) Liu, Y.; Wang, K.; Xu, X.; Eid, K.; Abdullah, A. M.; Pan, L.; Yamauchi, Y. Recent Advances in Faradic Electrochemical Deionization: System ArchitecturesversusElectrode Materials. *ACS Nano*. 2021. https://doi.org/10.1021/acsnano.1c03417.